\documentclass[prb,twocolumn,amsmath,amssymb,nopacs,superscriptaddress]{revtex4-1}
\usepackage[utf8]{inputenc}
\usepackage[dvips]{graphicx}
\usepackage{color}
\usepackage{latexsym}
\usepackage{amsmath}
\usepackage{amsfonts}
\usepackage{amssymb}
\usepackage{bm}
\usepackage{tabularx}

\newcommand{\fig}[2]{\includegraphics[width=#1]{#2}}
\newcommand{\bS}{{\bf S}}
\newcommand{\br}{{\bf r}}
\newcommand{\bQ}{{\bf Q}}
\newcommand{\bq}{{\bf q}}
\newcommand{\bk}{{\bf k}}
\newcommand{\bp}{{\bf p}}
\newcommand{\noi}{\noindent}

\newcolumntype{L}[1]{>{\raggedright\arraybackslash}p{#1}}
\newcolumntype{C}[1]{>{\centering\arraybackslash}p{#1}}
\newcolumntype{R}[1]{>{\raggedleft\arraybackslash}p{#1}}

\begin{document}

\title{Antiferromagnetic chiral spin density wave and strain-induced Chern insulator\\ in the square lattice Hubbard model with frustration}

\author{Yun-Peng Huang}
\affiliation{CAS Key Laboratory of Theoretical Physics, Institute of Theoretical Physics, Chinese Academy of Sciences, Beijing 100190, China}
\affiliation{School of Physical Sciences, University of Chinese Academy of Sciences, Beijing 100049, China}

\author{Jin-Wei Dong}
\affiliation{CAS Key Laboratory of Theoretical Physics, Institute of Theoretical Physics, Chinese Academy of Sciences, Beijing 100190, China}
\affiliation{School of Physical Sciences, University of Chinese Academy of Sciences, Beijing 100049, China}

\author{Panagiotis Kotetes}
\email{kotetes@itp.ac.cn}
\affiliation{CAS Key Laboratory of Theoretical Physics, Institute of Theoretical Physics, Chinese Academy of Sciences, Beijing 100190, China}

\author{Sen Zhou}
\email{zhousen@itp.ac.cn}
\affiliation{CAS Key Laboratory of Theoretical Physics, Institute of Theoretical Physics, Chinese Academy of Sciences, Beijing 100190, China}
\affiliation{School of Physical Sciences, University of Chinese Academy of Sciences, Beijing 100049, China}
\affiliation{CAS Center for Excellence in Topological Quantum Computation, University of Chinese Academy of Sciences, Beijing 100049, China}

\begin{abstract}
We employ the Hartree-Fock approximation to identify the magnetic ground state of the Hubbard model on a frustrated square lattice. We investigate the phase diagram as a function of the Coulomb repulsion's strength $U$, and the ratio $t'/t$ between the nearest and next nearest neighbor hoppings $t$ and $t'$. At half-filling and for a sufficiently large $U$, an antiferromagnetic chiral spin density wave order with nonzero spin chirality emerges as the ground state in a wide regime of the phase diagram near $t'/t=1/\sqrt{2}$, where the Fermi surface is well-nested for both $(\pi,\pi)$ and $(\pi,0)/(0,\pi)$ wave vectors. This triple-${\bf Q}$ chiral phase is sandwiched by a single-$\bQ$ N\'{e}el phase and a double-${\bf Q}$ coplanar spin-vortex crystal phase, at smaller and larger $t'/t$, respectively. The energy spectrum in the chiral spin density wave phase consists of four pairs of degenerate bands. These give rise to two pairs of Dirac cones with the same chirality at the point $({\pi \over 2},{\pi\over 2})$ of the  Brillouin zone. We demonstrate that the application of a diagonal strain induces a $d_{xy}$-wave next nearest neighbor hopping which, in turn, opens gaps in the two Dirac cones with opposite masses. As a result, four pairs of well-separated topologically-nontrivial bands emerge, and each pair of those contributes with a Chern number $\pm1$. At half-filling, this leads to a zero total Chern number and renders the topologically-notrivial properties observable only in the ac response regime. Instead, we show that at $3/4$ filling, the triple-${\bf Q}$ chiral phase yields a Chern insulator exhibiting the quantum anomalous Hall effect.
\end{abstract}

\maketitle

\section{Introduction \label{sec0}}

The chiral spin density wave ($\chi$SDW) has attracted much attention in condensed matter physics, as it is di\-stinct for the net spin chirality~\cite{Wen89} $\chi_{ijk}=\langle\bS_i\cdot(\bS_j\times\bS_k)\rangle\neq0$ that it threads through a triangular plaquette defined by three lattice sites $\bm{R}_{i,j,k}$. When itinerant electrons move under its influence, they feel a spontaneous gauge flux that leads to the accumulation of a nonzero Berry phase~\cite{Berry}, which in turn gives rise to an anomalous contribution to the Hall coefficient~\cite{Karplus1954,Ye1999,Ohgushi2000,Taguchi2001,Shindou2001,Nagaosa2010,Niu}. Remarkably, this already takes place in the absence of an external magnetic field. Even more, when the bulk energy spectrum is fully gapped, such an anomalous contribution takes quantized values as a result of the nonzero total Chern number~\cite{Thouless,TKNN,Niu85} ${\cal C}$ of the occupied bands. In this manner, it opens perspectives for a topologically-nontrivial Chern insulator, i.e. with ${\cal C}\neq0$, which leads to the quantum anomalous Hall effect~\cite{Ohgushi2000,Nagaosa2010,Niu}~(QAHE). Besides the QAHE, the breaking of both parity and time-reversal symmetries in the $\chi$SDW phase brings about a number of intriguing phenomena~\cite{WeinbergBook,VolovikBook,FradkinBook}, such as, the occurence of pa\-ri\-ty anomaly~\cite{NiemiSemenoff,Redlich,Semenoff,Haldane,Yakovenko90}, anyon superconductivity~\cite{Laughlin1988}, anomalous thermoelectricity~\cite{Niu}, and the polar Kerr effect~\cite{Bennett,Kapitulnik}, which constitute cha\-ra\-cte\-ristic features of systems belonging to the anomalous Hall metal and insulator classes~\cite{Raghu,Tewari2008,Kotetes2008,CZhang,Kotetes2010,Maciejko,YuScience,Chakravarty,Kotetes2014}.

Among the various candidates for a $\chi$SDW, the so-called antiferromagnetic (AFM) $\chi$SDW with zero net magnetization ($\sum_i\bS_i=\bm{0}$) is particularly interesting, since AFM spin couplings and magnetic orders are ubi\-qui\-tous in correlated electronic systems. In fact, the AFM $\chi$SDW order has been experimentally discovered in the NiS$_2$~\cite{Miyadai1975,Kikuchi1978,Kikuchi1978b,Matsuura2003} and FeMn~\cite{Endoh1971,Tajima1976,Kennedy1987,Kawarazaki1990} antiferromagnets on the frustrated face-centered-cubic (fcc) lattice. Neutron scattering experiments~\cite{Kikuchi1978,Matsuura2003,Miyadai1975,Endoh1971,Tajima1976,Kikuchi1978b} observed a noncoplanar AFM order with a four-sublattice structure and three magnetic or\-de\-ring wave vectors. Moreover, it was inferred that the ordered spin moments on the four sublattices form a tetrahedron in spin space. On the theo\-re\-ti\-cal side, such noncoplanar and chiral magnetic orders have been intensively explored in the context of the Kondo lattice model~\cite{Akagi2010,Akagi2012,Akagi2013,Barros2013,Kato2010,Martin2008,Ozawa2014,Hayami2015,Chern2010,Barros2014,Chern2014,Ghosh2016,Hayami2016,Agterberg2000}, the Hu\-bbard model~\cite{Martin2008,Venderbos2012,Ran2014,Jiang2015,Kiesel2012,Li2012,Wang2012}, and Heisenberg spin mo\-dels~\cite{Kumar2010,Ran2014,Venderbos2012} on va\-rious two-dimensional and three-dimensional lattice structures. Specifically, it has been suggested that an AFM $\chi$SDW order can be stabilized on the triangular~\cite{Martin2008,Kato2010,Akagi2010,Akagi2012,Barros2013,Rahmani2013,Akagi2013,Ozawa2014,Kumar2010,Venderbos2012,Chern2012,Hayami2016}, honeycomb~\cite{Kiesel2012,Wang2012,Ran2014,Li2012,Nandkishore2012,Jiang2015}, kagome~\cite{Chern2014,Barros2014,Ghosh2016}, pyrochlore~\cite{Chern2010}, cubic~\cite{Hayami2014}, and fcc~\cite{Yosida1981,Yoshimori1981,Hirai1985,Shindou2001,Agterberg2000} lattices. In these systems, the three ordering wave vectors of the $\chi$SDW phase are equivalent by means of the point group symmetry of the crystal. Moreover, these ordering vectors are half of the fundamental reciprocal lattice vectors of the system. In\-te\-re\-stin\-gly, numerical calculations found that an AFM $\chi$SDW can develop even in some decorated va\-riants of the square lattice, e.g. the checkerboard~\cite{Venderbos2012b} and square-to-triangular lattices~\cite{Hayami2015}, which do not support three equi\-va\-lent wave vectors. While a microscopic theo\-ry for the AFM $\chi$SDW is currently lacking, it is ge\-ne\-ral\-ly believed that both electron correlation and geometric frustration play important roles in its stabilization.

In this work, we explore the magnetic states and discuss the possible realization of the AFM $\chi$SDW phase in the Hubbard model on a frustrated square lattice, with the frustration introduced by considering both the nea\-rest neighbor (NN) hopping $t$ and the next nearest neighbor (NNN) hopping $t'$ in the kinetic energy part of the model. Motivated by the experimental findings, we here consider that the ordering wave vectors of the AFM $\chi$SDW are half of the fundamental reciprocal lattice vectors. Therefore, our general investigation is focused on magnetic states with ordering wave vectors $\bQ_1=(\pi,0)$, $\bQ_2=(0,\pi)$, and $\bQ_3=(\pi,\pi)$ on the square lattice with the above type of hopping-induced frustration.

At half-filling, we find that the desired triple-$\bQ$ AFM $\chi$SDW phase is the ground state in an extended regime of the $(U,t'/t)$ phase diagram. The resulting band dispersions exhibit two twofold-degenerate Dirac cones pos\-ses\-sing the same chirality. These are located at the N=$({\pi\over 2},{\pi\over 2})$ point of the Brillouin zone (BZ). Remarkably, in this case, the system is an insulator despite the pre\-sen\-ce of the two Dirac points where multiple band tou\-chings occur. This is because the Dirac points are split in energy and are found above and below the Fermi level while, at the same time, the ma\-xi\-mum bandwidth of the reconstructed bands is smaller than this ener\-gy splitting. Given this spontaneously de\-ve\-lo\-ped magnetic ground state and the resulting band structure, we propose a mechanism that gaps out the Dirac points, and thus renders the system a Chern insulator. As we discuss in detail, this is possible by considering the effect of strain along the diagonal direction. The latter introduces a $d_{xy}$-wave NNN hopping $\tilde{t}$, which in turn gaps out the Dirac points by inducing mass terms of opposite signs. As a result, each pair of degenerate bands contributes with a Chern number of $\pm 1$, which implies that the total Chern number at half-filling is zero and a QAHE is unobservable. However, other electron filling fractions can support the QAHE. We explicitly demonstrate that this is the case for a 3/4 filling factor.

Before proceeding with our main analysis, we remark that the present work is restricted to the interplay of magnetic instabilities generated by Fermi surface (FS) ne\-stings only at wave vectors $\bQ_{1,2,3}$. While in this manner possible magnetic instabilities at other wave vectors are neglected, we argue that our strategy is still valid and worthwhile to pursue. First of all, our approach is justified by the actual experimental observation of such a triple-$\bQ$ AFM order, and the fact that we here attempt a qualitative exploration of the possible magnetic orders that become accessible in such a setting. In the same spirit, the tight-binding model employed here mainly serves the purpose of investigating the desired triple-$\bQ$ degeneracy and interplay, and is not targeted to make a strong connection to the band structure of a specific material. Even more, the results and discussion presented in this work would not change qualitatively upon the modification of the tight-binding parameters, since one of our main goals is to highlight the magnetic phases which become accessible upon such a coexistence. Finally, restricting our study to these three wave vectors also appears as the natural assumption when approa\-ching the problem using a strong-coupling model with an AFM superexchange~\cite{SuperAnderson} coupling $J$, since the short-ranged nature of the coupling favors ordering at these wave vectors.

The presentation of our methods and results are unfolded in the remaining five sections. Section~\ref{sec1} introduces the Hubbard model considered throughout, and highlights the rich physics emerging from it, thus supporting our motivation to study this problem. This is achieved by exposing the FS nesting pro\-per\-ties of the band structure in the nonmagnetic phase and the resul\-ting behavior of the bare static spin susceptibility for various parameter values. In Sec.~\ref{sec2}, we derive the respective mean-field Hamiltonian by treating the local Coulomb interaction within the Hartree-Fock appro\-xi\-ma\-tion. In addition, we present the possible magnetic ground states within the restricted subspace of magnetic wave vectors. Section~\ref{sec3} contains our numerically obtained magnetic phase diagrams in the $(U,t'/t)$ pa\-ra\-me\-ter plane at half-filling, where we find a wide regime where the AFM $\chi$SDW insulator is the ground state. In Sec.~\ref{sec4}, we consider the situation of an electron filling fraction of 3/4, where a topologically-nontrivial Chern insulator featuring the QAHE is realized by introducing diagonal strain. Section~\ref{sec5} contains our conclusions.

\begin{figure}[t!]
\begin{center}
\fig{4.3in}{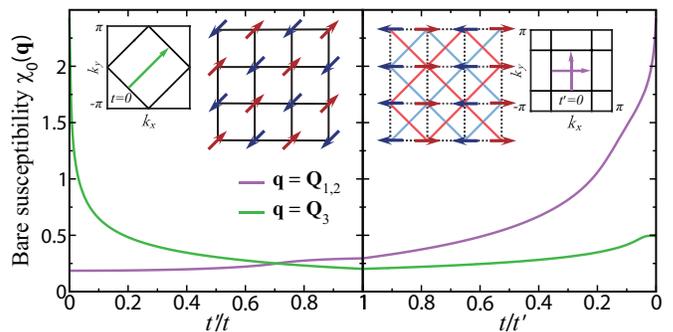}
\caption{(Color online) The static bare spin susceptibilities $\chi_0(\bq)$ at the three ordering wave vectors $\bQ_1=(\pi,0)$, $\bQ_2=(0,\pi)$, and $\bQ_3=(\pi,\pi)$ as functions of the ratio $t'/t$ or $t/t'$. The energy unit is set to be the NN (NNN) hopping $t=1$ ($t'=1$) in the left (right) panel. In the left (right) panel we have included two insets, which depict the Fermi surface topology appearing for $t'/t=0$ ($t/t'=0$), as well as the respective N\'eel (stripe) phase established. As the first inset of the right panel shows, the original square lattice is divisible into two enlarged square lattices which are illustrated by red and blue solid lines. In this manner, the $\bm{Q}_1$ stripe can be viewed as two antialigned N\'eel orders on the enlarged sublattices.}
\label{fig1}
\end{center}
\end{figure}

\section{Hubbard model and bare spin susceptibility analysis}\label{sec1}

We start with the Hubbard model on the square lattice, described by the following Hamiltonian
\begin{equation}
\mathcal{H}=-\sum_{i,j,\alpha}t_{ij}c^\dagger_{i\alpha}c_{j\alpha}+U\sum_i\hat{n}_{i\uparrow}\hat{n}_{i\downarrow},
\label{Hhub}
\end{equation}

\noi where $c^\dagger_{i\alpha} (c_{i\alpha})$ creates (annihilates) an electron with spin $\alpha=(\uparrow, \downarrow)$ at lattice site $i$, and $\hat{n}_{i\alpha} = c^\dagger_{i\alpha} c_{i\alpha}$ is the corresponding particle number operator. In the kinetic energy part, we consider both the NN and NNN hoppings to introduce frustration. Their combined presence leads to the following tight-binding energy dispersion
\begin{equation}
\epsilon_{\bf k}=-2t\left(\cos k_{x}+\cos k_{y}\right)-4t'\cos k_{x} \cos k_{y},\label{ek}
\end{equation}

\noi with the lattice constant set equal to unity and $t'$ considered to be positive throughout. Note that our sign choice for $t'$ becomes irrelevant at half-filling due to an emergent particle-hole symmetry.

The nesting properties of the FS resulting from $\epsilon_\bk$ play an important role in understan\-ding the structure of the long-range magnetic order that develops due to the pre\-sen\-ce of Coulomb repulsion of strength $U$, in the weak-strength limit. The emergence of nesting at the three wave vectors of interest $\bQ_{1,2,3}$ is reflected in the behavior of the bare static spin suscepti\-bi\-li\-ty $\chi_0(\bQ_{1,2,3})$. Fi\-gu\-re~\ref{fig1} depicts $\chi_0(\bQ_{1,2,3})$ as a function of the $t'/t$ or $t/t'$ ratio at half-filling, where the NN (NNN) hopping $t$ ($t'$) is set as the reference energy unit in the left (right) panel.

When we consider only the NN hopping $t$ at half-filling, \textit{i.e.}, $t'/t=0$, the corresponding FS is perfectly nested with $\bQ_3$. The presence of perfect nesting leads to a logarithmic divergence in $\chi_{0}(\bQ_3)$, as shown in Fig.~\ref{fig1}. Due to this divergence, a collinear magnetic order with ordering wave vector $\bQ_3$ develops even for an arbitrarily small $U$, and stabilizes the standard N\'{e}el phase.

In the inverse limit, where the NN hopping $t$ is zero, \textit{i.e.}, $t/t'=0$, the square lattice can be divided into two decoupled unfrustrated square lattices with a lattice constant enlarged by a factor of $\sqrt{2}$. The FSs for these two decoupled square lattices are perfectly nested with the wave vectors $(\pi,\pi)$ in the corresponding reduced Brillouin zones (RBZs). The latter wave vectors correspond to $\bQ_1$ and $\bQ_2$ in the original BZ. Consequently, two decoupled N\'{e}el AFM orders develop at any nonzero $U$ for both enlarged square lattices. Introducing a small amount of NN hopping $t$ couples the two N\'{e}el AFM orders, and the resulting FS does not show perfect nesting features any longer. As a result, a thre\-shold strength of $U$ is now required for the emergence of a magnetic order with ordering wave vector $\bQ_1$ and/or $\bQ_2$. More inte\-re\-stin\-gly, since the FS is now nested simultaneously by two wave vectors which are equivalent by virtue of the te\-tra\-go\-nal symmetry, a number of double-$\bQ$ magnetic orders become accessible~\cite{lorenzana08,eremin,brydon, giovannetti,gastiasoro15,wang15,kang15a,christensen15,Scherer2016,christensen17,christensen18,Fernandes2016}.

For the $\bQ_{1,2}$ wave vectors discussed here, there exist two possible double-$\bQ$ phases\cite{lorenzana08}, \textit{i.e.}, a collinear charge- and spin-ordered density wave (CSDW) phase, and a coplanar so-called spin-vortex crystal (SVC) phase where the moments on neighboring sites are at right angles to each other. Evidences for both of these phases have been recently experimentally recorded in Fe-based materials~\cite{hassinger,avci14a,wasser15,bohmer15a,allred15a,wang16a,zheng16a,malletta,mallettb,allred16a,meier17,Wang2019}. The tetragonal symmetry of the nonmagnetic phase further implies that the Stoner criteria for the single- and double-$\bQ$ phases are satisfied simultaneously. Thus, from a Landau theo\-ry perspective, both kinds of magnetic orders are de\-ge\-ne\-ra\-te at the quadratic level of the free-energy expansion in terms of the magnetic order pa\-ra\-me\-ter. All de\-ge\-ne\-ra\-cies are however lifted when considering the fourth-order contributions to the free ener\-gy~\cite{Agterberg2000,Hayami2014,lorenzana08,christensen18}.

When the frustration is strong, \textit{i.e.}, $t \sim t'$, the bare static spin susceptibilities $\chi_0 (\bq)$ at $\bQ_{1,2}$ and $\bQ_3$ become comparable as shown in Fig.~\ref{fig1}. In particular, their va\-lues are exactly the same at $t'/t = 1/\sqrt{2}\simeq 0.71$ where a Lifshitz~\cite{Lifshitz} transition modifies the FS topology. This implies that the FS is simultaneously nested by the three wave vectors $\bQ_{1,2,3}$, although these three wave vectors are not equivalent by means of the square lattice symmetry.

The diversity of magnetic order scenarios revealed from the above susceptibility analysis further supports our motivation to study here the interplay between phases ori\-gi\-na\-ting from the FS nesting with the three wave vectors, and explore the possible emergence of magnetic phases beyond the well-discussed single-$\bQ$ collinear stripe, and double-$\bQ$ CSDW and SVC orders~\cite{lorenzana08,eremin,brydon,giovannetti,gastiasoro15,wang15,kang15a,christensen15,Scherer2016,christensen17,Fernandes2016}. Notably, when the magnetic moments order at the three vectors simultaneously, the long-sought-after AFM $\chi$SDW phase becomes accessible in the fru\-stra\-ted regime of the present model, and opens perspectives for realizing a topologically-nontrivial AFM Chern insulator exhibiting the QAHE.

\section{Mean-field theory\label{sec2}}

To study the ground state properties of the Hubbard model, the Coulomb repulsion term in Eq.~\eqref{Hhub} is treated within the Hartree-Fock approximation which preserves the SU(2) spin-rotational symmetry of the interaction. The resulting mean-field Hamiltonian reads
\begin{align}
\mathcal{H}_\text{HF} =&-\sum_{i,j,\alpha} t_{ij} c^\dagger_{i\alpha}c_{j\alpha}+{U\over 4} \sum_i \big( 2 n_i \hat{n}_i -n^2_i \big) \nonumber \\
&- {U\over 4} \sum_i \big( 2{\bm m}_i \cdot \hat{\bm m}_i-{\bm m}^2_i\big),\label{Hmf}
\end{align}

\noi where ${\bm \sigma}=(\sigma_x,\sigma_y,\sigma_z)$ defines the vector of the Pauli matrices. In the above, $n_i$ and ${\bm m}_i$ denote the mean fields of the local particle density $\hat{n}_i=\sum_\alpha \hat{n}_{i\alpha}$ and magnetic moment $\hat{\bm m}_i=\sum_{\alpha\beta} c^\dagger_{i\alpha}{\bm \sigma}_{\alpha\beta} c_{i\beta}$ operators, respectively. These are obtained from the statistical ave\-ra\-ge of the corresponding operators with respect to the single-particle mean-field Hamiltonian in Eq.~\eqref{Hmf}, and are ge\-ne\-ral\-ly expressed as $n_i={\bar n}+\sum_\eta\widetilde{\mathcal{N}}_\eta \cos (\bQ_\eta \cdot {\bf r}_i + \theta_\eta )$ and ${\bm m}_i = \sum_\eta \widetilde{\bm M}_\eta \cos (\bQ_\eta \cdot {\bf r}_i + \theta'_\eta )$, with $\widetilde{\mathcal{N}}_\eta~(\widetilde{\bm M}_\eta )$ the charge~(magnetic) order parameters with ordering wave vector $\bQ_\eta$ and $\theta_\eta~(\theta'_\eta)$ the relative phases. $\bar n$ is the ave\-ra\-ge particle density per site which equals to $1$ at half-filling.

When the allowed ordering wave vectors $\bQ_\eta$ are limited to ($\pi$, 0), (0, $\pi$), and ($\pi$, $\pi$), and the lattice symmetry is tetragonal, the expressions simplify to
\begin{equation}
\{n_i,{\bm m}_i\} = \{\bar{n},\bm{0}\}+\sum_\eta\{\mathcal{N}_\eta,{\bm M}_\eta\}\cos{({\bf Q}_\eta\cdot\br_i)},
\end{equation}

\noi with the order parameters $\mathcal{N}_\eta = \widetilde{\mathcal{N}}_\eta \cos \theta_\eta$ and ${\bm M}_\eta = \widetilde{\bm M}_\eta\cos\theta'_\eta$. As a result, there exist four inequivalent lattice sites in the ordered phase, which lead to a $2\times 2$ enlarged unit cell. On the four inequivalent sites, the local particle density is $n_i= \{\bar n +\mathcal{N}_1 +\mathcal{N}_2 +\mathcal{N}_3, \bar n -\mathcal{N}_1 +\mathcal{N}_2 -\mathcal{N}_3, \bar n -\mathcal{N}_1 -\mathcal{N}_2 +\mathcal{N}_3, \bar n +\mathcal{N}_1 -\mathcal{N}_2 -\mathcal{N}_3\}$, and the local magnetic moment reads ${\bm m}_i =\{{\bm M}_1 +{\bm M}_2 +{\bm M}_3, -{\bm M}_1 +{\bm M}_2 -{\bm M}_3, -{\bm M}_1 -{\bm M}_2 +{\bm M}_3, {\bm M}_1 -{\bm M}_2 -{\bm M}_3\}$. It is important to note that the energy contribution per site from the charge order is $U\sum_\eta \mathcal{N}^2_\eta$. Therefore, any type of charge order is energe\-ti\-cal\-ly costly and disfavored and, as a result, the ground state of the Hubbard model is expected to be a state of a uniform charge density with all $\mathcal{N}_\eta=0$. For the magnetic order, the fourth-order expansion of the magnetic free energy, which we discuss in Appendix~\ref{app1}, shows that if multi-$\bQ$ ordering takes place, the ordered moments of the dif\-fe\-rent order parameters develop in a pairwise paral\-lel or per\-pen\-di\-cu\-lar fashion. This is because only such confi\-gu\-ra\-tions minimize the free energy.

When two or three magnetic order parameters have moments which are aligned in parallel, they give rise to local moments that have different amplitudes $|{\bm m}_i|$ on the four inequivalent sites. Thus, this induces some sort of charge order and leads to a CSDW-type of phase, which is not a likely ground state of the Hubbard model in Eq.~\eqref{Hhub}. This is due to the unavoidable energy penalty for de\-ve\-lo\-ping the charge order. Note, however, that there e\-xists numerical evidence for the CSDW phase in extended Hubbard mo\-dels with additional interactions that not con\-si\-de\-red here~\cite{lorenzana08,eremin,brydon, giovannetti,gastiasoro15,wang15,kang15a,christensen15,Scherer2016,christensen17,christensen18,Fernandes2016}, as well as experimental proof that this phase is realized in certain Fe-based compounds~\cite{allred16a}. The above corroborate that the ground state of the Hubbard model in Eq.~\eqref{Hhub} should be a phase with uniform charge and ordered magnetic moments which, when they arise, they align per\-pen\-di\-cu\-lar to each other. In fact, this feature is confirmed by our unrestricted numerical calculations. Hence, we hereinafter restrict our discussion to this kind of states with uniform charge density and ordered moments on the four sublattices. By virtue of the global SO(3) spin rotational symmetry of the model, we further fix the directions of ${\bm M}_1$, ${\bm M}_2$, and ${\bm M}_3$ to be along the $x$, $y$, and $z$ spin axis, respectively. Thus, without loss of generality, the ordered moments read ${\bm M}_1=(M_1,0,0)$, ${\bm M}_2=(0,M_2,0)$, and ${\bm M}_3=(0,0,M_3)$, and yield ${\bm m}_1=(M_1,M_2,M_3)$, ${\bm m}_2=(-M_1,M_2,-M_3)$, ${\bm m}_3=(-M_1,-M_2,M_3)$, and ${\bm m}_4=(M_1,-M_2,-M_3)$.

The commensurate character of the magnetic wave vectors allows us to more conveniently treat the problem in ${\bf k}$ (wave vector) space. Specifically, since ${\bf k}+2\bQ_{1,2}={\bf k}$ and $\bQ_3=\bQ_1+\bQ_2$, we find $\bQ_\eta$=$-\bQ_\eta$ and $\pm\bQ_1\pm\bQ_2\pm\bQ_3=\bm{0}$. The above relations hold modulo a shift by a reciprocal lattice vector, and imply that the ${\bf k}$-space mean-field Hamiltonian of Eq.~\eqref{Hmf} becomes
\begin{align}
\mathcal{H}_\text{HF}= & \sum_{{\bf k},\alpha} \epsilon_{\bf k} c_{\bf k \alpha}^{\dagger} c_{\bf k\alpha} -{U \over 2} \sum_{{\bf k},\alpha,\beta,\eta} M_\eta c^\dagger_{\bf k \alpha} \sigma^\eta_{\alpha\beta} c_{{\bf k}+{\bf Q}_{\eta},\beta} \nonumber \\
+& {1\over 4}NU\sum_\eta M^2_\eta\,.
\label{Hk}
\end{align}

\noi Interestingly, since the here-considered magnetic order parameters are invariant under a lattice translation combined with a spin rotation~\cite{Martin2008}, the Hamiltonian in Eq.~\eqref{Hk} can be split into the following two identical disjoint parts
\begin{equation}
\mathcal{H}_\text{HF}=\sum_{\bk\in {\rm RBZ}}\left({\bm \Psi}^\dagger_\bk\hat{H}_\bk{\bm\Psi}_\bk+{\bm\Phi}^\dagger_\bk\hat{H}_\bk{\bm\Phi}_\bk\right),\label{Hmatrix}
\end{equation}

\noi with the spinors ${\bm \Psi}_\bk$=$(c_{\bk \uparrow}$, $c_{\bk+\bQ_1 \downarrow}$, $c_{\bk+\bQ_2 \downarrow}$, $c_{\bk+\bQ_3 \uparrow})^\intercal$, ${\bm \Phi}_\bk$=$(c_{\bk \downarrow}$, $c_{\bk+\bQ_1 \uparrow}$, $-c_{\bk+\bQ_2 \uparrow}$, $-c_{\bk+\bQ_3 \downarrow})^\intercal$, and the $\bk$-dependent $4\times 4$ Hamiltonian matrix
\begin{equation}
\hat{H}_\bk=\left(\begin{array}{cccc}
\epsilon_\bk & -{1\over 2} UM_1  & {i\over 2}UM_2 & -{1\over 2} UM_3\\
-{1\over 2}UM_1 & \epsilon_{\bk+\bQ_1} & {1\over 2}UM_3 & -{i\over 2}UM_2 \\
-{i\over 2}UM_2 & {1\over 2}UM_3 & \epsilon_{\bk+\bQ_2} & -{1\over 2}UM_1\\
-{1\over 2} UM_3 & {i\over 2}UM_2 & -{1\over 2}UM_1 & \epsilon_{\bk+\bQ_3}
\end{array}\right). \label{matrix}
\end{equation}

\noi As a result, the reconstructed band structure consists of four pairs of degenerate bands. Note that the wave vector summation in Eq.~\eqref{Hmatrix} is over the RBZ, which corresponds to one quarter of the original BZ, due to the four-sublattice structure of the direct lattice.

To investigate the ground state properties, we mi\-ni\-mi\-ze the energy by obtaining the magnetic order pa\-ra\-me\-ter associated with each ordering wave vector $\bQ_\eta$ self-consistently at zero temperature, via the relations
\begin{equation}
M_\eta={1\over N}\sum_{\bk,\alpha,\beta} \left\langle c^\dagger_{\bk \alpha} \sigma^\eta_{\alpha\beta}c_{\bk+\bQ_\eta,\beta}\right\rangle,
\end{equation}

\noi with $N$ denoting the number of $\bk$-points in the RBZ. When a magnetic order at a single $\bQ$ develops, it gives rise to a collinear state which is termed as the N\'{e}el or the stripe phase, depending on whether the ordering vector is $\bQ_3$ or $\bQ_{1,2}$. The $\bQ_{1,2}$ stripe phases are equivalent by means of the fourfold rotational symmetry of the energy dispersion. The coplanar SVC magnetic phase carrying a vector chi\-ra\-li\-ty $\bm{\chi}_{ij}=\langle\bS_i\times\bS_j\rangle\neq\bm{0}$ is achieved when two magnetic orders emerge at the same time. All coplanar phases obtained in this work exhibit ordering at $\bQ_1$ and $\bQ_2$, and have moments of equal amplitude which lie in the $xy$ spin plane. When all three magnetic orders coexist simultaneously, the AFM $\chi$SDW phase is realized with the spin-chirality value $\chi={1\over 2} M_1 M_2 M_3$ per unit cell. Note that the tetragonal point group symmetry enforces $|M_1|=|M_2|$ in the $\chi$SDW phase.

In summary, besides the paramagnetic (PM) metal phase with all $M_\eta=0$, which is obtained at small strengths of Coulomb repulsion, the minimization of the energy leads to four possible magnetic ground states: (a) the N\'{e}el phase with $(M_1,M_2,M_3)=(0,0,M)$, (b) the stripe phase with $(M_1,M_2,M_3)=(M,0,0)$ or, equivalently, $(0,M,0)$, (c) the SVC phase with $(M_1,M_2,M_3)=(M,M,0)$, and (d) the $\chi$SDW with $(M_1,M_2,M_3)=(M,M,M')$.

Finally, we additionally note that, for our numerical simulations, we employ different initial conditions for sol\-ving the self-consistency equations for a given choice of the set of Hamiltonian parameters. When different ground states are obtained for the different initial conditions, we compare the state energies of the different configurations in order to infer the true ground state.

\section{AFM $\bm \chi$SDW at half-filling \label{sec3}}

We first explore the magnetic orders of the frustrated square lattice Hubbard model at half-filling with $\bar n=1$. Note that our study goes beyond the exploration of previous related works~\cite{Yu2010,Zou2012,Yamada2013,Mizusaki2006,Nevidomskyy2008,Tocchio2008}, which did not consider the possibility of the $\chi$SDW phase.

\subsection{Phase diagram and phase transitions}

\begin{figure}[t!]
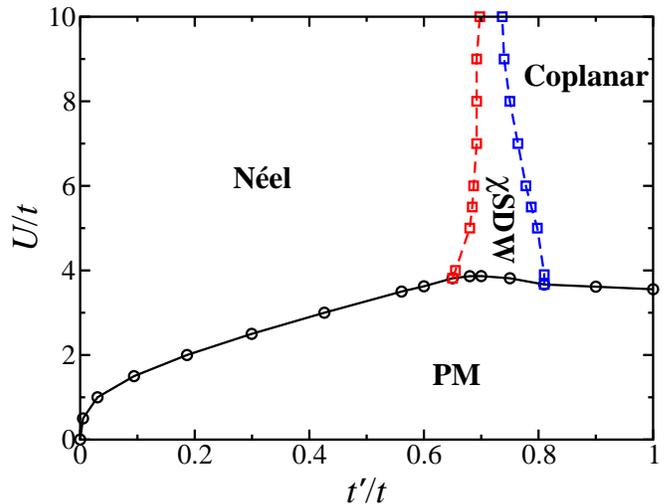

\begin{center}
\fig{3.4in}{fig2}
\caption{(Color online) Phase diagram of the half-filled square lattice Hubbard model with frustration. Solid and dashed lines denote, respectively, phase boundaries of first-order and continuous transitions.}
\label{fig2}
\end{center}
\end{figure}

The magnetic phase diagram at half-filling is presented in Fig.~\ref{fig2} as a function of the Hubbard repulsion strength $U$ and the $t'/t$ ratio. Hereafter, we take the strength of the NN hopping $t$ as the ener\-gy unit. The possible ground states of the phase diagram are spanned by the PM metal, single-$\bQ$ N\'{e}el, double-$\bQ$ coplanar SVC phase, and triple-$\bQ$ $\chi$SDW phases. The boundaries between these phases are determined by comparing the state energies of different phases. The solid and dashed lines denote, respectively, a first-order and a continuous phase transition between two neighboring phases. When $U$ is not sufficiently strong to stabilize any long-range magnetic order, the ground state is a PM metal with all $M_\eta = 0$.

Magnetic ordering generally emerges only above a cri\-ti\-cal $U_c$, and the precise structure of the magnetic order is governed by the $t'/t$ ratio. At small $t'/t$, the nesting at wave vector $\bQ_3$ is much stronger than that at $\bQ_{1,2}$, and leads to the N\'{e}el phase. Note that an infinitesimally weak $U$ is capable of dri\-ving a N\'{e}el phase at $t'/t=0$, due to the divergence of the spin susceptibility at $\bQ_3$. As the $t'/t$ ratio increases, a higher critical $U_c$ is required. The situation is different at large $t'/t$, where the nesting is stronger at $\bQ_{1,2}$. Our numerical calculations reveal that the ground state is the double-$\bQ$ coplanar SVC phase which takes advantage of the nestings at both wave vectors. The $\bQ_1$ or $\bQ_2$ stripe phases reside higher in energy. The critical $U_c$ decreases as $t'/t$ increases, and is expected to reach zero in the limit of $t'/t \rightarrow \infty$ (\textit{i.e.}, $t/t' = 0$), where the susceptibility $\chi_0(\bQ_{1,2})$ diverges. Remarkably, in a significantly wide regime about $t'/t=1/\sqrt{2}$, where the nestings at $\bQ_{1,2}$ and $\bQ_3$ are comparable in strength, the AFM $\chi$SDW emerges as the ground state. All three ordered moments develop simultaneously to lower the state energy, giving rise to a nonzero spin chirality which opens perspectives for an anomalous Hall response.

\begin{figure}[t!]
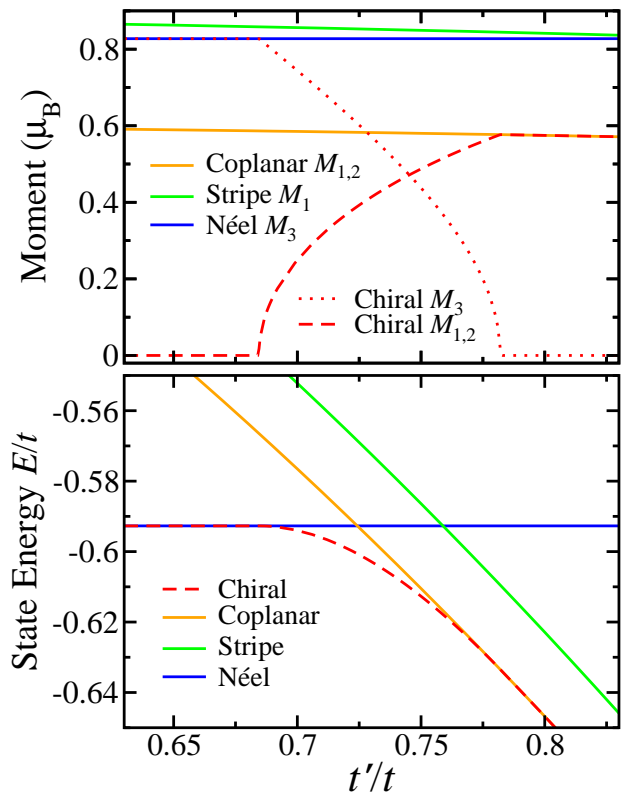

\begin{center}
\fig{3.2in}{fig3}
\caption{(Color online) (a) The ordered magnetic moments and (b) the state energy per site of the four distinct magnetic phases as a function of the $t'/t$ ratio at half-filling when the strength of the Coulomb repulsion is $U=6t$.}
\label{fig3}
\end{center}
\end{figure}

All magnetic ground states presented in Fig.~\ref{fig2} are insulating and, as $U$ decreases, they give their place to a PM metal phase via first-order transitions. To investigate the phase transitions between the different magnetic ground states, we consider $U=6t$, and monitor the evolution of the four distinct magnetic phases of interest as a function of the $t'/t$ ratio, with a focus on the transitions. The results are summarized in Fig.~\ref{fig3}. The ordered magnetic moments for the four magnetic phases are plotted in Fig.~\ref{fig3}(a), and their energies per site are compared in Fig.~\ref{fig3}(b). The amplitude of the ordered moments in the four phases show only a weak dependence on the frustration ratio $t'/t$, since $U$ is quite strong in the regime displayed in Fig.~\ref{fig3}.

We find that the N\'{e}el phase is lower in energy than the stripe and coplanar phases at small $t'/t$. Note that although the ordered moment has a slightly larger amplitude in the stripe phase (see for instance $M_1$ in the $\bQ_1$-ordered stripe phase) than in the coplanar phase ($\sqrt{M^2_1+M^2_2}$), the stripe phase is always higher in ener\-gy than the coplanar SVC phase, as the latter utilizes both ordering wave vectors. As $t'/t$ increases, the inplane magnetic orders $M_1$ and $M_2$ become favored, and emerge at $t'/t \simeq 0.69$, where the out-of-plane magnetic order $M_3$ starts to decrease, thus reflecting the competition between the FS nestings at wave vectors $\bQ_{1,2}$ and $\bQ_3$. A further increase of $t'/t$, leads to the complete suppression of $M_3$ at $t'/t\simeq 0.78$, and the ground state converges to the double-$\bQ$ coplanar SVC phase. The AFM $\chi$SDW phase is obtained in the regime of $0.69\lesssim t'/t \lesssim 0.78$, where all three magnetic orders coexist. When approa\-ching its phase boundaries, the energy of the AFM $\chi$SDW gradually becomes equal to that of the N\'{e}el and coplanar SVC phases, as shown in Fig.~\ref{fig3}(b). The transitions between the three magnetic phases in the phase diagram are therefore continuous, with the boundaries denoted by the dashed lines in Fig.~\ref{fig2}.

\subsection{Band dispersion and Dirac cones}

\begin{figure}[t!]
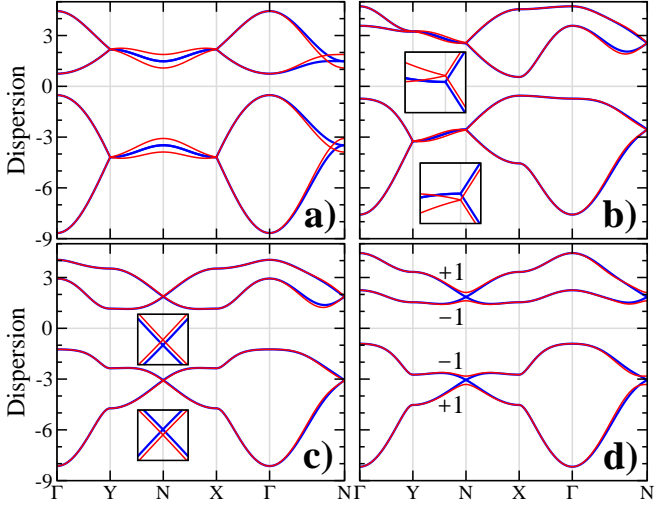

\begin{center}
\fig{3.4in}{fig4}
\caption{(Color online) Band dispersions in the four magnetically-ordered phases, (a) N\'{e}el, (b) stripe, (c) coplanar SVC, and (d) $\chi$SDW, along the high-symmetry path in the RBZ at half-filling and $(U,t')=(6, 0.74)t$, where the $\chi$SDW is the ground state and the other three metastable states. $\Gamma=(0,0)$, $\text{Y}=(0,{\pi\over2})$, $\text{X}=({\pi\over2},0)$, and $\text{N}=({\pi\over 2}, {\pi\over 2})$. The blue and red lines denote, respectively, the band dispersion with the $d_{xy}$-wave NNN hopping $\tilde{t}=0$ and $\tilde{t}=0.1t$. The numbers in (d) denote the Chern numbers of each pair of bands with $\tilde{t}=0.1t$. Insets in (b) and (c) are enlargements of the band dispersions near the N point.}
\label{fig4}
\end{center}
\end{figure}

The band dispersions are readily obtained by diago\-na\-li\-zing the matrix Hamiltonian of Eq.~\eqref{matrix}. The spectrum of each magnetic phases has four pairs of twofold degenerate bands, with two of them above and the other two below the Fermi energy. For a direct comparison of the band dispersions arising in the four dis\-tinct magnetic phases, we consider the parameter values $(U,t')=(6,0.74)t$, and depict the resulting energy dispersions in Fig.~\ref{fig4}. For these pa\-ra\-me\-ter va\-lues, the AFM $\chi$SDW constitutes the ground state, while the remaining three can be viewed as metastable phases correspon\-ding to local minima of the free energy. To obtain the displayed band structures for the metastable pha\-ses, we use the self-consistently obtained magnetic moments by restricting our evaluation to the vicinity of the local mi\-ni\-ma. Remarkably, in the AFM $\chi$SDW and coplanar SVC phases, we find two pairs of Dirac cones at the $\bk$-space point N=$({\pi\over2},{\pi\over2})$ as shown in Figs.~\ref{fig4}(c) and~\ref{fig4}(d).

The appearance of the Dirac cones in the coplanar SVC and AFM $\chi$SDW phases is much more evident in a rotated basis, with the unitary transformation matrix explicitly given in Appendix~\ref{app2}. In the rotated basis, the Hamiltonian $\hat{H}_\bk$ of Eq.~\eqref{matrix} becomes
\begin{equation}
\tilde{H}_\bk = \left(\begin{array}{cc} H_+ & H_w \\ H^\dagger_w & H_- \end{array}\right), \label{matrix2}
\end{equation}

\noi with the $\bk$-dependent $2\times 2$ matrices $H_\pm$ and $H_w$ correspondingly given by
\begin{align}
H_{\pm}= & \pm \big(\Gamma_0 \sigma_0 -\Gamma' \cos \theta \sigma_z \big) \nonumber \\
&+\Gamma_x \sin\theta \sin\varphi \sigma_x -\Gamma_y \sin\theta \cos\varphi \sigma_y, \label{Hpm} \\
H_{w}= & -\Gamma' \sin\theta \sigma_0 -\Gamma_x \big( \cos\varphi -i\cos\theta \sin\varphi \big) \sigma_y \nonumber\\
& +\Gamma_y \big( \sin\varphi +i\cos\theta \cos\varphi \big) \sigma_x. \label{Hw}
\end{align}

\noi Here $\Gamma_0=\frac{1}{2} U M$ , $\Gamma_x =2t\cos k_x$, $\Gamma_y= 2t\cos k_y$, $\Gamma' =4t' \cos k_x \cos k_y$, and $M$, $\theta$, $\varphi$ are, respectively, the radial distance, polar angle, and azimuthal angle in the spherical coordinate of the magnetic moment on the first sublattice site ${\bm m}_1=(M_1,M_2,M_3)$. At the N point, where $\cos k_x=\cos k_y=0$, $\Gamma_x=\Gamma_y=\Gamma'=0$, the Hamiltonian matrix $\tilde{H}_\bk$ becomes diagonal with two twofold degenerate eigenvalues $\pm\Gamma_0$. We expand the Hamiltonian around the N point and set $\bk=({\pi\over 2},{\pi\over 2})+\bp$. At leading order in $\bp$, we find:
\begin{align}
H_\pm=&\pm\Gamma_0\sigma_0+2t\sin\theta\big(\sin\varphi p_x\sigma_x-\cos\varphi p_y\sigma_y\big),\label{Hpm2} \\
H_w=&-2t\big(\cos\varphi-i\cos\theta\sin\varphi\big)p_x\sigma_y \nonumber\\
&+2t\big(\sin\varphi+i\cos\theta \cos\varphi\big) p_y \sigma_x.\label{Hw2}
\end{align}

\noi Note that neither the diagonal nor the off-diagonal blocks contain $\Gamma'$, as this appears only at quadratic order in $\bp$.

In the AFM $\chi$SDW phase, where $\sin\theta$ is nonzero and $\varphi ={\pi\over 4}$, the diagonal blocks $H_\pm$ give rise to two iso\-tropic Dirac cones at energies $\pm \Gamma_0$, with the same chirality, and velocity of value $\sqrt{2}t\sin\theta$. Even more, the off-diagonal blocks $H_w$ and $H^\dagger_w$, are linear in $\bp$ and va\-nish at the N point, thus preserving the Dirac cones. In a similar fashion, the spectrum of the coplanar SVC phase with $\theta={\pi\over2}$ and $\varphi={\pi\over4}$ also exhibits two Dirac cones at N, as shown in Fig.~\ref{fig4}(c). In the single-$\bQ$ stripe phase, take $\bQ_1$-ordered phase for example, $\theta={\pi\over 2}$ and $\varphi=0$. The diagonal blocks thus become $H_\pm=\pm\Gamma_0\sigma_0-2t p_y\sigma_y$, that does not lift the twofold degeneracy along the N-Y RBZ line, where $p_y=0$ (\textit{i.e.}, $k_y = {\pi\over 2}$), and fails to ge\-ne\-ra\-te Dirac cones in the spectrum. Taking into account the off-diagonal blocks $H_w=-2tp_x\sigma_y$, the dispersions of the four pairs of bands in the stripe phase are given by $\pm \sqrt{\Gamma^2_0 +4t^2 p^2_x } \pm 2t p_y$, producing the spectrum shown in Fig.~\ref{fig4}(b). In the $\bQ_3$-ordered N\'{e}el phase, $\theta=0$, and thus the diagonal and off-diagonal blocks read $H_\pm=\pm\Gamma_0\sigma_0$, $H_w=2te^{i\varphi}(p_y\sigma_x+ip_x\sigma_y)$, therefore lea\-ding to the band dispersion $\pm\sqrt{\Gamma^2_0+4t^2(p_x\pm p_y)^2}$. The bands are fourfold degenerate along both N-X and N-Y directions where, either $p_x$ or $p_y$ equals zero, as shown in Fig.~\ref{fig4}(a).

\subsection{Strain-induced gap opening and band topology}

To lift the additional degeneracy at the N point and thus gap out the Dirac cones, we consider the pre\-sen\-ce of strain applied in the diagonal direction, which breaks the fourfold rotational symmetry. This vio\-la\-tion introduces a $d_{xy}$-wave hopping $\tilde{t}$ on the NNN bonds. When $\tilde{t}=t'$ ($\tilde{t}=t'=t/2$), the lattice effectively becomes square-to-triangular~\cite{Hayami2015} (triangular~\cite{Martin2008}), and the spectrum becomes fully-gapped in the AFM $\chi$SDW phase. The $d_{xy}$-wave NNN hopping modifies the tight-binding dispersion $\epsilon_\bk$ in Eq.~\eqref{ek}, since its effect is reflected in the addition of the term $4\tilde{t}\sin k_x\sin k_y$. Consequently, the term $\Gamma'$ of Eqs.~\eqref{Hpm}-\eqref{Hw} acquires an extra contribution of $-4\tilde{t}\sin k_x\sin k_y$, which is of the order of $1$ near the N point, and is no longer negligible when the Hamiltonian is expanded in terms of $\bp$. Indeed, the diagonal blocks $H_\pm$ of Eq.~\eqref{Hpm2} and the off-diagonal blocks $H_w$ of Eq.~\eqref{Hw2} receive, respectively, the additional term $\pm 4\tilde{t} \cos\theta \sigma_z$ and $-4\tilde{t}\sin\theta\sigma_0$. As a result, in the triple-$\bQ$ chiral phase, where $\cos\theta\neq 0$, two mass terms with opposite masses $\pm 4\tilde{t}\cos\theta$ are introduced to the two Dirac cones stemming from the diagonal blocks. The Dirac cones are therefore gapped out, as shown in Fig.~\ref{fig4}(d), where the AFM $\chi$SDW phase is obtained self-consistently in the presence of $d_{xy}$-wave NNN hopping $\tilde{t}$. On the other hand, the diagonal blocks $H_\pm$ describing the Dirac cones are unaltered in the double-$\bQ$ coplanar SVC phase with $\theta={\pi\over 2}$. Speci\-fi\-cal\-ly, despite the fact that the term $-4\tilde{t}\sigma_0$ is added to the off-diagonal blocks, it leads to a mere shift of the energies of the two Dirac cones, from the initial $\pm\Gamma_0$ to the final $\pm\sqrt{\Gamma^2_0+16\tilde{t}^2}$ energies.

The full gap that is induced in the $\chi$SDW phase due to addition of the $d_{xy}$-wave NNN hopping, generally gives rise to a nonzero total Chern number ${\cal C}$, and a concomitant QAHE at zero temperature with Hall conductance:
\begin{equation}
\sigma_{xy}=-{e^2\over h}{\cal C}\quad{\rm where}\quad{\cal C}=\sum_s^{\rm occupied}\int\frac{d^2k}{2\pi}\Omega_{s\bk}\,.
\end{equation}

\noi In the above, $\Omega_{s\bk}=i\varepsilon_{zij}\big<\partial_{k_i}\bm{u}_{s\bk}\big|\partial_{k_j}\bm{u}_{s\bk}\big>$ ($i,j=x,y$) denotes the Berry curvature~\cite{Niu} of the $s$-th occupied quasiparticle reconstructed band with eigenvector $\big|\bm{u}_{s\bk}\big>$. Note that we employed the Einstein summation convention, introduced the totally-antisymmetric symbol $\varepsilon_{zij}$, and converted the summation to an integration by considering the continuum $N\rightarrow\infty$ limit.

We find that the strain-induced mass term has an opposite sign on the two Dirac cones at the N point. This implies that each pair of degenerate bands, from top to bottom, contributes with a Chern number of $\{+1,-1,-1,+1\}$, as indicated in Fig.~\ref{fig4}(d). Although the dc Hall conductance is zero at half-filling, as the two pairs of occupied bands have opposite Chern numbers, the Chern bands shown in Fig.~\ref{fig4}(d) still allow for an anomalous Hall effect in the ac regime and the emergence of anomalous optical dichroism~\cite{Jiang2015} which becomes accessible via interband transitions.

\section{Strain-induced Chern insulator at 3/4 filling \label{sec4}}

\begin{figure}[t!]
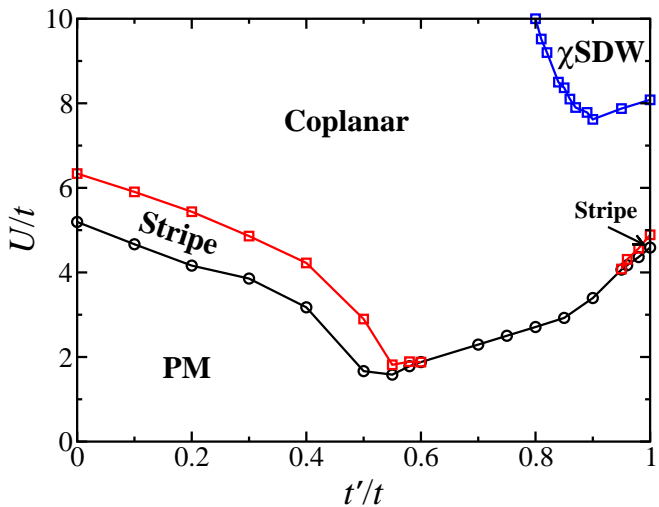

\begin{center}
\fig{3.4in}{fig5}
\caption{(Color online) Phase diagram of the frustrated square lattice Hubbard model at 3/4 filling. Additional third NN hopping with $t''=0.4t'$ is introduced in the model to stabilize the AFM $\chi$SDW as a ground state in the phase diagram.}
\label{fig5}
\end{center}
\end{figure}

In order to obtain an AFM Chern insulator which exhibits the QAHE at zero temperature, it is required to modify the occupation of the bands, so that only an odd number of these are filled. We find that this becomes possible when doping the system away from half-filling to 3/4 filling. We note that the AFM $\chi$SDW phase does not emerge as a ground state of the 3/4-filled Hubbard model when only considering the NN and NNN hoppings. However, adding a third NN hopping $t''$ stabilizes it. Therefore, in this section, we introduce a $t''$ and fix it to the value $t''=0.4 t'$. The third NN hopping $t''$ modifies the tight-binding dispersion $\epsilon_\bk$ by introducing an extra term $-\Gamma''$, with $\Gamma''=2t''(\cos 2k_x$ $+\cos 2k_y)$. Consequently, the diagonal blocks $H_\pm$, receive an additional term of $\Gamma'' \sigma_0$ in the rotated basis, which shifts the energies of the Dirac cones but does not lift any degeneracy.

\begin{figure}[t!]
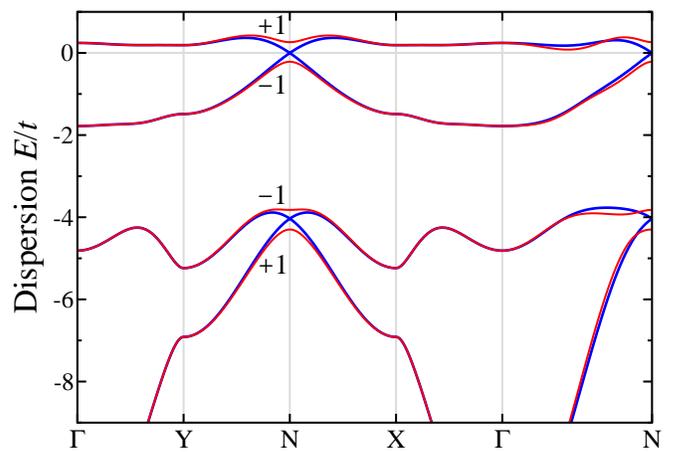

\begin{center}
\vspace{0.1cm}
\fig{3.4in}{fig6}
\caption{(Color online) Band dispersion in the $\chi$SDW phase along a high-symmetry path in the RBZ at 3/4 filling and $(U, t')=(9,0.84)t$. The blue and red lines denote the band dispersion with the $d_{xy}$-wave NNN hopping $\tilde{t}=0$ and $\tilde{t}=0.1 t$, respectively. The numbers in the figure denote the Chern numbers of each pair of bands with $\tilde{t}=0.1t$.}
\label{fig6}
\end{center}
\end{figure}

In Fig.~\ref{fig5} we present the $(U,t'/t)$ phase diagram at 3/4 filling. All phase boundaries correspond to critical lines of first-order transitions. The ground state is a PM metal at small $U$, while upon its increase, a coplanar SVC phase is established. In particular, the coplanar SVC phase directly succeeds the PM phase in the window $0.6< t'/t<0.95$, while for $t'/t < 0.6$ and $t'/t > 0.95$, an intermediate stripe phase appears. The triple-$\bQ$ $\chi$SDW phase is stabilized in the upper-right corner of the phase diagram, where $t'\sim t$ and $U$ are significant. Notably, due to the fourfold degeneracy at the N point, the magnetic ordered phases are metallic at 3/4 filling. In particular, the double-$\bQ$ coplanar phase and the triple-$\bQ$ $\chi$SDW phase give rise to a Dirac semimetal. See also Fig.~\ref{fig6}.

As put forward in the previous section, the Dirac cones at the N point can be gapped out by applying strain along the diagonal direction, since this brings about a $d_{xy}$-wave NNN hopping. The band dispersion of the $\chi$SDW phase with $\tilde{t}=0.1t$ is shown in Fig.~\ref{fig6}. Indeed, the Dirac cones are gapped out and all four pairs of bands are well-separated from each other. Each pair of bands, from top to bottom, contributes with a Chern number $\{+1, -1, -1, +1\}$. Remar\-ka\-bly, since only the lower three pairs of bands are occupied at 3/4 filling, the total Chern number is $-1$, giving rise to a topologically-nontrivial Chern insulator which features a QAHE at zero tem\-pe\-ra\-tu\-re.

\section{Conclusions \label{sec5}}

In this work, we investigate the magnetic orders and phase transitions which arise in the square lattice Hubbard model with frustration, where the allowed ordering wave vectors are restricted to $\bQ_1 =(\pi,0)$, $\bQ_2 =(0,\pi)$, and $\bQ_3 =(\pi,\pi)$. To study the ground state properties, the Hartree-Fock approximation is applied to the local Coulomb interaction. When the strength of the interaction is sufficiently strong, the ground state at half-filling is a $\bQ_3$-ordered N\'{e}el phase at small $t'/t$, and a double-$\bQ$ coplanar SVC phase at large $t'/t$. Interestingly, an AFM $\chi$SDW phase is stabilized in a wide regime of the phase diagram near $t'/t=1/\sqrt{2}$, where the nestings of the Fermi surface at wave vectors $\bQ_{1,2}$ and $\bQ_3$ are comparable in strength. Here, the three magnetic orders coexist to utilize simultaneously the nestings at the three wave vectors, and give rise to a noncoplanar magnetic phase with a nonzero spin chirality. The phase transitions from the N\'{e}el to the $\chi$SDW and from the $\chi$SDW to the coplanar SVC phase are continuous.

We find that the energy spectrum of the $\chi$SDW phase contains two pairs of Dirac cones, which are located at the Brillouin zone point N=$({\pi \over 2},{\pi \over 2})$ and possess the same chirality. We show that applying strain along the dia\-go\-nal direction, introduces a $d_{xy}$-wave next-nearest neighbor hopping, which in turn gaps out the two Dirac cones with opposite masses. This gives rise to four pairs of well-separated topologically-nontrivial bands. Each pair of bands contributes with a Chern number $\pm1$, and the total Chern number is zero at half-filling. Finally, we show that doping the system to a 3/4 filling with a nonzero third nearest neighbor hopping, stabilizes the $\chi$SDW phase and leads to a topologically-nontrivial Chern insulator which features the quantum anomalous Hall effect.

Concluding this work, we wish to make a number of remarks regarding the validity of the Hartree-Fock mean-field approximation, as well as the stability of the magnetic order and the resulting Chern insulator state. First of all, we need to stress that the $\chi$SDW phase was found here for interaction strengths $U/t$ larger than $4$ or $7$ depending on the filling which, while they are certainly not weak, they still remain sufficiently smaller than the value required to be reached for transiting to a Mott insulator phase. Thus, having stayed safely away from the Mott transition further justifies our assumption of the pre\-sen\-ce of quasiparticles, as well as considering a well-defined Fermi surface as our starting point. In fact, in the pa\-ra\-me\-ter regime of inte\-rest, the ratio of the hopping strengths sa\-ti\-sfies $t'/t\sim1$, thus implying a bandwidth of about $12t$. Hence, we expect the impact of the repulsive interaction to be effectively much weaker than it would be for the standard Hubbard model with $t'=0$.

The validity of our Hartree-Fock approximation and the correctness of this simple intuitive picture is further supported by a number of checks that we have carried out using the SU(2) spin-rotation invariant slave-boson method~\cite{SBpaper1,SBpaper2,SBpaper3}. Employing this method allows us to explore the magnetic phase diagram in the strong-coupling limit. The calculations presented in Appendix~\ref{app3} show that the two methods provide qualitatively similar results. This further backs the validity of the weak-coupling approach employed here, and further guarantees the stability of the topological phe\-no\-me\-na proposed in this work against quantum fluctuations. In fact, even in their presence, the topological properties remain unaltered, given that the gap in the bulk spectrum remains intact and the bands preserve their topologically-nontrivial character.

While providing a precise estimate of the degree of topological robustness in the presence of fluctuations is beyond the scope of this work, we assert that these become irrelevant as long as the modification of the order parameters they induce is much smaller than the magnitudes of the order parameters obtained at the mean-field level. Notably, going beyond mean-field theory and properly incorporating fluctuations requires accounting for the re\-mai\-ning spatial dimension of the magnetic system, which has been neglected here throughout, as this is customary in the study of strongly anisotropic systems with a quasi-2D behaviour. In fact, the stabilization of long range magnetic order in the present work impli\-ci\-tly assumes the third dimension, since this is crucial for circumventing the Hohenberg-Mermin-Wagner theorem.

\section{Acknowledgments}

This work is supported by the CAS Key Research Program of Frontier Sciences (Grant No.~QYZDB-SSW-SYS012), the Strategic Priority Research Program of CAS (Grant No.~XDB28000000), and the National Na\-tu\-ral Science Foundation of China (Grant No.~11747601 and~11974362). Numerical calculations were performed on the HPC Cluster of ITP-CAS.

\appendix

\section{Landau-type analysis of the magnetic free energy} \label{app1}

To perform a Landau-type magnetic instability ana\-ly\-sis for the model under consideration, we obtain the free energy density up to quartic order in terms of the magnetic order parameters ${\bm M}_{1,2,3}$ with ordering wave vector $\bQ_{1,2,3}$, which reads:
\begin{align}
F & =\alpha \left( \bm{M}_1^2+\bm{M}_2^2+\bm{M}_3^2\right)+\beta\left(\bm{M}_1^4+\bm{M}_2^4+\bm{M}_3^4\right)\nonumber \\
 & +\gamma \left( \bm{M}_1^2 \bm{M}_2^2 +\bm{M}_1^2 \bm{M}_3^2 +\bm{M}_2^2  \bm{M}_3^2 \right) \nonumber \\
 & +\eta \left[ \left( \bm{M}_1 \cdot \bm{M}_2 \right)^2 +\left( \bm{M}_1 \cdot \bm{M}_3 \right)^2 +\left( \bm{M}_2 \cdot \bm{M}_3 \right)^2 \right] \nonumber \\
& +\delta \alpha \bm{M}_3^2 +\delta \beta \bm{M}_3^4  +\delta \gamma \bm{M}_1^2 \bm{M}_2^2  +\delta \eta \left( \bm{M}_1 \cdot \bm{M}_2 \right)^{2}. \nonumber
\end{align}

\noi The anisotropic terms in the last line of the above expression are present to account for the fact that $\bQ_3$ is ine\-qui\-va\-lent to $\bQ_{1,2}$ given a square lattice symmetry. The Landau expansion contains all symmetry-allowed terms, and therefore enables the discussion of all possible magnetic instabilities, independently of the underlying microscopic mechanism.

We proceed by parametrizing the magne\-tic order pa\-ra\-me\-ters by $\bm{M}_1 = M \sin \theta \cos \varphi \hat{\bm n}_1$, $\bm{M}_2 = M \sin \theta \sin \varphi \hat{\bm n}_2$, and $\bm{M}_3 = M \cos \theta \hat{\bm n}_3$, with $M= (\bm{M}^2_1 +\bm{M}^2_2 +\bm{M}^2_3 )^{1/2}$, and the angles  $\theta \in [0, \pi]$, $\varphi \in [0, 2\pi)$.
The unit vectors along the direction of $\bm{M}_{1,2,3}$ are denoted as $\hat{\bm n}_{1,2,3}$, with the angle $\phi_{ij}$ between $\hat{\bm n}_i$ and $\hat{\bm n}_j$ given by $\cos \phi_{ij} =\hat{\bm n}_i \cdot \hat{\bm n}_j$. In terms of these newly defined parameters, the Landau free energy becomes
\begin{align}
F& =(\alpha +\delta\alpha \cos^2 \theta ) M^2 \nonumber \\
&+\left[ \beta \sin^4 \theta \left( \cos^4 \varphi +\sin^4 \varphi \right) +\left( \beta +\delta \beta \right) \cos^4\theta\right] M^4 \nonumber \\
&+{1\over 4}\left[ (\gamma+\delta \gamma) \sin^4 \theta \sin^2 2\varphi + \gamma \sin^2 2\theta \right] M^4 \nonumber \\
&+{1\over 4} \big[ (\eta +\delta \eta)\sin^4 \theta\sin^2 2\varphi \cos^2 \phi_{12} \nonumber\\
&+\eta \sin^2 2\theta (\cos^2 \varphi \cos^2 \phi_{13} + \sin^2 \varphi \cos^2\phi_{23})\big] M^4. \nonumber
\end{align}

\noi Minimizing the Landau free energy with respect to the angles $\phi_{12}$, $\phi_{13}$, and $\phi_{23}$, yields the following three corresponding equations of motion
\begin{align}
\sin^4 \theta \sin^2(2\varphi)\sin(2\phi_{12})&=0,\nonumber \\
\sin^2 (2\theta)\cos^2\varphi\sin(2\phi_{13})&=0,\nonumber \\
\sin^2 (2\theta)\sin^2\varphi\sin(2\phi_{23})&=0.\nonumber
\end{align}

\noi The above equations are satisfied only when the $\phi$ angles are multiples of ${\pi \over 2}$. The latter implies that the ordered magnetic moments, when developed, are pairwise parallel or perpendicular to each other.\\

\section{Transformation of the Hamiltonian\label{app2}}

For generality, we consider here the case with the third NN hopping $t''$ and the strain-induced $d_{xy}$-wave NN hopping $\tilde{t}$. The tight-binding dispersion entering the Hamiltonian matrix of Eq.~\eqref{matrix} becomes: $\epsilon_\bk = -\Gamma_x -\Gamma_y -\Gamma' -\Gamma'', $ with $\Gamma_x =2t\cos k_x$, $\Gamma_y =2t\cos k_y$, $\Gamma' =4t' \cos k_x \cos k_y$ $- 4\tilde{t} \sin k_x \sin k_y$, and $\Gamma'' = 2t''(\cos 2k_x + \cos 2k_y)$. Parametrizing the magnetic order parameters by $M_1 = M \sin \theta \cos \varphi$, $M_2 = M \sin \theta \sin \varphi$, and $M_3 = M \cos \theta$, with $M= (M^2_1 +M^2_2 +M^2_3 )^{1/2}$, the Hamiltonian matrix can be rewritten as $H=H_t+H_m$ with:
\begin{align}
H_{t}=&-\Gamma_x \tau_0 \otimes \sigma_z -\Gamma_y \tau_z \otimes \sigma_0 -\Gamma' \tau_z \otimes \sigma_z -\Gamma'' \tau_0 \otimes \sigma_0,\nonumber\\
H_{m}=&-\Gamma_0 \sin \theta \cos \varphi \tau_0 \otimes \sigma_x -\Gamma_0 \sin \theta \sin \varphi \tau_y \otimes \sigma_z \nonumber\\
&+\Gamma_0 \cos \theta \tau_y \otimes \sigma_y,\nonumber
\end{align}

\noi where $\Gamma_0 = {1\over 2} U M$, and $\sigma_{0,x,y,z}$, $\tau_{0,x,y,z}$ are the $2\times2$ identity matrices and Pauli matrices. Under a unitary transformation of
$U=U_1e^{i{\varphi\over2}\tau_z\otimes\sigma_z} e^{i {\theta \over 2}\tau_z\otimes\sigma_y}U_2,$ with
\begin{equation}
U_1= {1\over 2}\left(\begin{array}{cccc}
1 & -i & -i & -1\\
-i & 1 & -1 & -i\\
i & 1 & -1 & i\\
1 & i & i & -1 \end{array}\right),\,
U_2= {1\over \sqrt{2}}\left(\begin{array}{cccc}
0 & 0 & 1 & -1\\
1 & 1 & 0 & 0\\
1 & -1 & 0 & 0\\
0 & 0 & 1 & 1 \end{array}\right)\nonumber
\end{equation}

\noi the magnetic Hamiltonian becomes diagonal, \textit{i.e.}, $H_m= \Gamma_0 \tau_z \otimes \sigma_0$, and the kinetic Hamiltonian $H_t$ becomes
\begin{align}
H_{t} & =\Gamma_x \sin \theta \sin \varphi \tau_0 \otimes \sigma_x -\Gamma_y \sin \theta \cos \varphi \tau_0 \otimes \sigma_y\nonumber\\
& -\Gamma' \cos \theta \tau_z \otimes \sigma_z -\Gamma' \sin \theta \tau_x \otimes \sigma_0 \nonumber\\
& -\Gamma_x \cos \theta \sin \varphi \tau_y \otimes \sigma_y -\Gamma_x \cos \varphi \tau_x \otimes \sigma_y \nonumber\\
& -\Gamma_y \cos \theta \cos \varphi \tau_y \otimes \sigma_x +\Gamma_y \sin \varphi \tau_x \otimes \sigma_x-\Gamma'' \tau_0 \otimes \sigma_0. \nonumber
\end{align}

\noi Sorting into $2\times 2$ diagonal and off-diagonal blocks, the Hamiltonian matrix in Eq.~\eqref{matrix} reads in the rotated basis,
\begin{equation}
\tilde{H} = \left(\begin{array}{cc} H_+ & H_w \\ H^\dagger_w & H_- \end{array}\right), \nonumber
\end{equation}

\noi with the wave-vector dependent $2\times 2$ blocks $H_\pm$ and $H_w$ given by, respectively,
\begin{align}
H_{\pm}= & \pm \left(\Gamma_0 \sigma_0 -\Gamma' \cos \theta \sigma_z \right) -\Gamma'' \sigma_0 \nonumber \\
&+\Gamma_x \sin\theta \sin\varphi \sigma_x -\Gamma_y \sin\theta \cos\varphi \sigma_y, \nonumber \\
H_{w}= & -\Gamma' \sin\theta \sigma_0 -\Gamma_x \left( \cos\varphi -i\cos\theta \sin\varphi \right) \sigma_y \nonumber\\
& \hspace{1.85cm} +\Gamma_y \left( \sin\varphi +i\cos\theta \cos\varphi \right) \sigma_x. \nonumber
\end{align}

\section{Slave-boson method calculations\label{app3}}

In this section we present additional calculations based on the SU(2) spin-rotation-invariant slave-boson (SRISB) method~\cite{SBpaper1,SBpaper2,SBpaper3}, that we carry out in order to test the validity of the weak-coupling mean-field approach adopted in the main text. Since the SRISB method has been already discussed in the literature extensively, here, we only briefly mention its essential ingredients. Instead, we focus on the main assumptions considered to study the Hamiltonian in Eq.~\ref{Hhub}, and discuss our findings.

As it is customary in the SRISB approach, we express the original fermionic creation operators as $c_{i\alpha}\equiv\sum_\beta f_{i\beta}z_{i\alpha\beta}$, with $f_{i\alpha}$ defining fermionic operators and $z_{i\alpha\beta}$ being functions of the bosonic operators $e_i$, $p_{i0}$, $\bm{p}_i=\left(p_{ix},p_{iy},p_{iz}\right)$ and $d_i$.

To faithfully replace the original fermionic operators, the following constraints are imposed per lattice site $i$:
\begin{eqnarray}
d_i^\dag d_i+e_i^\dag e_i+\sum_{\mu=0,x,y,z} p_{i\mu}^\dag p_{i\mu}-1&=&0\,,\qquad\\
\sum_{\alpha=\uparrow,\downarrow} f_{i\alpha}^{\dagger}f_{i\alpha}-\sum_{\mu=0,x,y,z}p_{i\mu}^\dag p_{i\mu}-2d_i^\dag d_i&=&0\,,\qquad\\
\sum_{\alpha,\beta=\uparrow,\downarrow}f_{i\alpha}^{\dagger}\bm{\sigma}_{\alpha\beta}f_{i\beta}-p_{i0}^\dag\bm{p}_i-\bm{p}_i^\dag p_{i0}-i\bm{p}_i^\dag\times\bm{p}_i&=&0\,.
\end{eqnarray}

\noi The above constraints are enforced by introducing the Lagrange multipliers $\alpha_{i}$, $\beta_{i,0}$ and $\bm{\beta}_i$, respectively. Within the SRISB framework, the Hubbard interaction term takes the quadratic form $\sim d_i^\dag d_i$, while the electron hopping becomes dependent on the various bosonic fields.

\begin{figure}[t!]
\centering
\includegraphics[width=1\columnwidth]{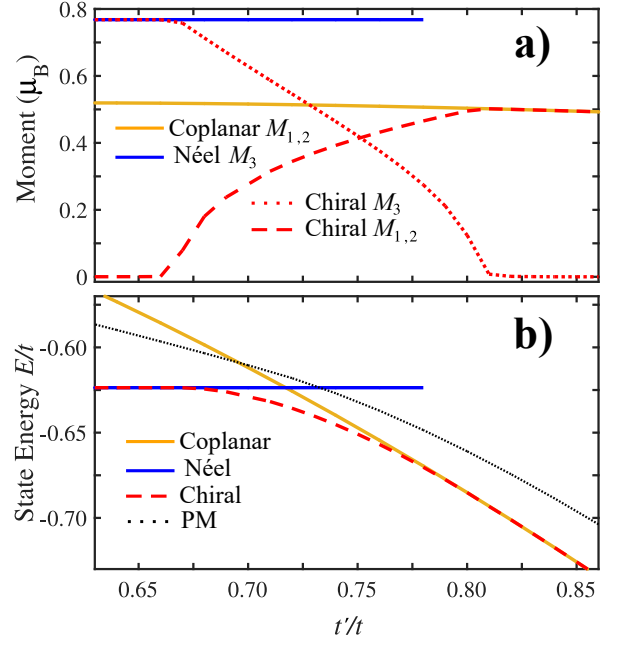}
\caption{SU(2) spin-rotation-invariant slave-boson mean-field calculations at half-filling, when the strength of the Coulomb repulsion is $U=6t$. (a) The magnetic moments in the various magnetic states, and (b) the corresponding state energy per site as a function of the $t^{\prime}/t$ ratio. Note that, as in the main text, the magnetic moments for states which do not constitute the ground state, are obtained by restricting our evaluation to the vicinity of the respective local mi\-ni\-mum.}
\label{fig:app3a}
\end{figure}

We proceed by mean-field decoupling the bosonic fields using the spatiotemporally-independent ans\"atze: $e_i\mapsto e\in\mathbb{R}^+$, $p_{i0}\mapsto p_0\in\mathbb{R}^{+}$, $d_i\mapsto d\in\mathbb{R}^+$, $\beta_{i0}\mapsto\beta_0\in\mathbb{R}$, $\alpha_i\mapsto\alpha\in\mathbb{R}$, $p_{i0}\mapsto p_0\in\mathbb{R}^+$, $d_i\mapsto d\in\mathbb{R}^{+}$, $\beta_{i0}\mapsto\beta_0\in\mathbb{R}$, $\alpha_i\mapsto\alpha\in\mathbb{R}$, $\bm{p}_i\mapsto p\bm{g}$, $\bm{\beta}_i\mapsto\beta\bm{g}$, with $ p\in\mathbb{R}^+$, $\beta\in\mathbb{R}$, and $\bm{g}=\big(\sin\theta\cos\varphi\cos(\bm{{\rm Q}}_1\cdot\bm{{\rm r}}_i),\sin\theta\sin\varphi\cos(\bm{{\rm Q}}_2\cdot\bm{{\rm r}}_i), \cos\theta\cos(\bm{{\rm Q}}_3\cdot\bm{{\rm r}}_i)\big)$.

Under these ans\"atze, the matrix functions $z_i$ can be written as $z_i=Z_{+}\sigma_0+Z_{-}\sum_{\mu=0,x,y,z}p_{\mu}\sigma_\mu/|\bm{p}|$, where we introduced $Z_{\pm}=(z_+\pm z_-)/2$ with:
\begin{equation}
z_{\pm}=\frac{\big[p_{0}\left(e+d\right)\pm p\left(e-d\right)\big]/\sqrt{2}}{\sqrt{\left[1-\left|d\right|^{2}-\left(p_{0}\pm p\right)^{2}/2\right]\left[1-\left|e\right|^{2}-\left(p_{0}\mp p\right)^{2}/2\right]}}.
\end{equation}

\begin{figure}[t!]
\centering
\includegraphics[width=1\columnwidth]{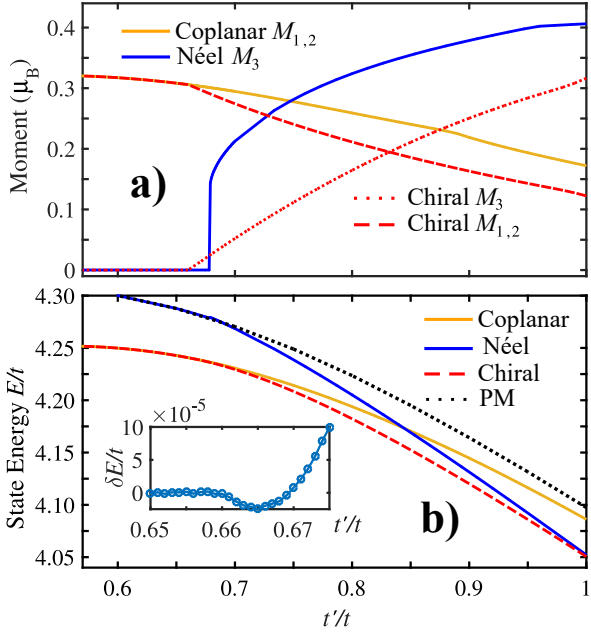}
\caption{SU(2) spin-rotation-invariant slave-boson mean-field calculations at $3/4$-filling, when the strength of the Coulomb repulsion is $U=10t$, and the ratio of $t^{\prime\prime}/t^{\prime}=0.4$. (a) The magnetic moments in the various magnetic states, and (b) the corresponding state energy per site. The inset in (b) shows the energy difference of the chiral and SVC phases $\delta E$, obtained as a function of the $t^{\prime}/t$ ratio. Note that, as in the main text, the magnetic moments for states which do not constitute the ground state, are obtained by restricting our evaluation to the vicinity of the respective local mi\-ni\-mum.}
\label{fig:app3b}
\end{figure}

Given the above definitions, we choose a basis for the $f$ fermions which is analogous to the one defined in Eq.~\eqref{Hmatrix}, i.e., $\tilde{\bm{\Psi}}_{\bm{{\rm k}}}^\dag=(f_{\bm{{\rm k}}\uparrow},f_{\bm{{\rm k}}+\bm{{\rm Q}}_1\downarrow},f_{\bm{{\rm k}}+\bm{{\rm Q}}_2\downarrow},f_{\bm{{\rm k}}+\bm{{\rm Q}}_3\uparrow})$
and $\tilde{\bm{\Phi}}_{\bm{{\rm k}}}^\dag=(f_{\bm{{\rm k}}\downarrow},f_{\bm{{\rm k}}+\bm{{\rm Q}}_1\uparrow},-f_{\bm{{\rm k}}+\bm{{\rm Q}}_2\uparrow},-f_{\bm{{\rm k}}+\bm{{\rm Q}}_3\downarrow})$. This choice allows decomposing the $f$-fermion Hamiltonian into two identical parts, i.e., $\mathcal{H}_\text{SRISB}^f=\sum_{\bk\in {\rm RBZ}}(\tilde{{\bm \Psi}}^\dagger_\bk\hat{H}_\bk^f\tilde{{\bm\Psi}}_\bk+\tilde{{\bm\Phi}}^\dagger_\bk\hat{H}_{\bm{{\rm k}}}^f\tilde{{\bm\Phi}}_\bk)$. We find that $\hat{H}_\bk^f=\hat{H}_\bk'+\delta\hat{H}_\bk$. The first term has an identical structure to $\hat{H}_{\bm{{\rm k}}}$ in Eq.~\eqref{Hmatrix} but with renor\-ma\-li\-zed coefficients and reads:
\begin{eqnarray}
\hat{H}_{\bm{{\rm k}}}'=&-&\left(A_1-A_2\right)\Gamma_x\tau_0\otimes\sigma_z-\left(A_1+A_2\right)\Gamma_y\tau_z\otimes\sigma_0\nonumber\\
&-&B_1\Gamma'\tau_z\otimes\sigma_z-B_2\Gamma''\tau_0\otimes\sigma_0-\beta\sin\theta\cos\varphi\tau_0\otimes\sigma_x\nonumber\\
&-&\beta\sin\theta\sin\varphi\tau_y\otimes\sigma_z+\beta\cos\theta\tau_y\otimes\sigma_y\,.
\end{eqnarray}

\noi In contrast, the second term contains matrix elements nonpreexisting in $\hat{H}_\bk$, i.e.:
\begin{eqnarray}
\delta\hat{H}_\bk=&-&C_1\Gamma_y\tau_z\otimes\sigma_x-C_1\Gamma''\tau_0\otimes\sigma_x+C_2\Gamma_x\tau_0\otimes\sigma_y\nonumber\\
&-&C_2\Gamma'\tau_z\otimes\sigma_y-C_3\Gamma'\tau_x\otimes\sigma_x+C_3\Gamma''\tau_y\otimes\sigma_y\nonumber\\
&+&C_4\Gamma_y\tau_x\otimes\sigma_y-C_4\Gamma_x\tau_y\otimes\sigma_x-C_5\Gamma_x\tau_y\otimes\sigma_0\nonumber\\
&-&C_5\Gamma''\tau_y\otimes\sigma_z+C_6\Gamma_y\tau_x\otimes\sigma_z-C_6\Gamma'\tau_x\otimes\sigma_0\qquad\,\,
\end{eqnarray}

\noi where we defined the parameters: $A_1=Z_+^{2}-Z_-^2\cos^2\theta$, $A_2=Z_-^2\sin^2\theta\cos(2\varphi)$, $B_1=Z_+^2+Z_-^2\cos(2\theta)$, $B_2=Z_+^2+Z_-^2$
$C_1=2Z_+Z_-\sin\theta\cos\varphi$, $C_2=Z_-^2\sin(2\theta)\sin\varphi$, $C_3=2Z_+Z_-\cos\theta$, $C_4=Z_-^2\sin^2\theta\sin(2\varphi)$, $C_5=2Z_+Z_-\sin\theta\sin\varphi$, and $C_6=Z_-^2\sin2\theta\cos\varphi$.

Taking into account the contribution of the $f$-fermion Hamiltonian to the free energy, as well as the terms originating from the constraints and the Hubbard interaction, we obtain the intensive SRISB mean-field free energy:
\begin{eqnarray}
{\cal F}&=&-\frac{T}{N}\sum_{\bm{{\rm k}},s}\ln\Big[1+e^{-\left(\epsilon_{s\bm{{\rm k}}}^f-\mu_0+\beta_0\right)/T}\Big]+Ud^2\nonumber\\
&+&\mu_0n-2\beta p_{0}p-\beta_{0}\left(p_{0}^{2}+p^{2}+2d^{2}\right)\nonumber\\
&+&\alpha\left(e^{2}+p_{0}^{2}+d^{2}-1+p^{2}\right),\quad
\end{eqnarray}

\noindent where the $\epsilon_{s\mathrm{\boldsymbol{k}}}^f$ denotes the eigenenergies of $\hat{H}_\bk^f$ labelled by the band index $s$. The ground state of the system is found by minimizing the above free energy with respect to the set of variables $\{\mu_0,e,p_0,p,d,\alpha,\beta_0,\beta\}$.

The numerically-obtained magnetic phase diagram for $U=6t$ upon varying the ratio $t^{\prime}/t$ at half-filling is shown in Fig.~\ref{fig:app3a}. Similar to the weak-coupling calculation of the main text, also here, the system transits from the N\'{e}el phase to the chiral magnetic phase, and finally to the coplanar SVC phase as we increase $t'/t$. However, here, the chiral SDW state occupies a larger area of the phase diagram, i.e., $0.66\lesssim t'/t\lesssim0.81$, compared to the weak-coupling result, where we found $0.68\lesssim t'/t\lesssim0.78$.

Our numerical results for the $3/4$ filling are shown in Fig.~\ref{fig:app3b}. The system transits from the SVC to the chiral magnetic phase upon increasing $t'/t$. The corresponding transition is first order, thus agreeing with our main text result. Indeed, as shown in the inset of Fig.~\ref{fig:app3b}(b), the SVC and chiral SDW phases lead to the same free energy for $0.66\lesssim t'/t\lesssim0.67$.

\end{document}